\documentclass[sigconf]{acmart}
\AtBeginDocument{%
  }

\setcopyright{acmlicensed}
\copyrightyear{2026}
\acmYear{2026}
\setcopyright{cc}
\setcctype{by}
\acmConference[WWW '26]{Proceedings of the ACM Web Conference 2026}{April 13--17, 2026}{Dubai, United Arab Emirates}
\acmBooktitle{Proceedings of the ACM Web Conference 2026 (WWW '26), April 13--17, 2026, Dubai, United Arab Emirates}
\acmPrice{}
\acmDOI{10.1145/3774904.3792792}
\acmISBN{979-8-4007-2307-0/2026/04}

\usepackage{enumitem}
\usepackage{colortbl}  
\usepackage{xcolor}
\usepackage{url}
\usepackage{multirow}
\usepackage{subcaption}
\usepackage{arydshln}
\usepackage{tabularx}
\usepackage{enumitem}
\usepackage{array} 
\usepackage{balance}
\usepackage{booktabs}
\usepackage{graphicx}
\usepackage{algorithm}
\usepackage{algpseudocode}
\usepackage{float}
\usepackage{amsmath}
\usepackage{bm}
\usepackage{threeparttable}
\usepackage{xcolor}
\usepackage{dcolumn}
\definecolor{lightblue}{RGB}{235,245,255}

\let\oldhat\hat

\renewcommand{\hat}[1]{\oldhat{\mathbf{#1}}}

\begin{document}

\title{PI2I: A Personalized Item-Based Collaborative Filtering Retrieval Framework}

\author{Shaoqing Wang}
\authornote{Equal contribution}
\affiliation{%
\institution{Alibaba Group}
\city{Hangzhou}
\country{China}
}
\email{wangshaoqing.wsq@alibaba-inc.com}

\author{Yingcai Ma}
\authornotemark[1]
\affiliation{%
\institution{Alibaba Group}
\city{Hangzhou}
\country{China}
}
\email{yingcai.myc@alibaba-inc.com}

\author{Kairui Fu}
\authornotemark[1]
\authornote{Work is done during the internship at Alibaba Group.}
\affiliation{%
  \institution{Zhejiang University}
  \city{Hangzhou}
  \country{China}
}
\email{fukairui.fkr@zju.edu.cn}

\author{Ziyang Wang}
\affiliation{%
\institution{Alibaba Group}
\city{Hangzhou}
\country{China}
}
\email{shanyi.wzy@alibaba-inc.com}

\author{Dunxian Huang}
\affiliation{%
\institution{Alibaba Group}
\city{Hangzhou}
\country{China}
}
\email{dunxian.hdx@alibaba-inc.com}

\author{Yuliang Yan}
\affiliation{%
\institution{Alibaba Group}
\city{Hangzhou}
\country{China}
}
\email{yuliang.yyl@alibaba-inc.com}

\author{Jian Wu}
\affiliation{%
\institution{Alibaba Group}
\city{Beijing}
\country{China}
}
\email{joshuawu.wujian@alibaba-inc.com}

\renewcommand{\shortauthors}{Shaoqing Wang et al.}

\begin{abstract}
Efficiently selecting relevant content from vast candidate pools is a critical challenge in modern recommender systems. Traditional methods, such as item-to-item collaborative filtering (CF) and two-tower models, often fall short in capturing the complex user-item interactions due to uniform truncation strategies and overdue user-item crossing. To address these limitations, we propose \textbf{P}ersonalized \textbf{I}tem-to-\textbf{I}tem (\textbf{PI2I}), a novel two-stage retrieval framework that enhances the personalization capabilities of CF. In the first \textbf{I}ndexer \textbf{B}uilding \textbf{S}tage (IBS), we optimize the retrieval pool by relaxing truncation thresholds to maximize Hit Rate, thereby temporarily retaining more items users might be interested in. In the second \textbf{P}ersonalized \textbf{R}etrieval \textbf{S}tage (PRS), we introduce an interactive scoring model to overcome the limitations of inner product calculations, allowing for richer modeling of intricate user-item interactions. Additionally, we construct negative samples based on the trigger-target (item-to-item) relationship, ensuring consistency between offline training and online inference. Offline experiments on large-scale real-world datasets demonstrate that PI2I outperforms traditional CF methods and rivals Two-Tower models. Deployed in the "Guess You Like" section on Taobao, PI2I achieved a 1.05\% increase in online transaction rates. In addition, we have released a large-scale recommendation dataset collected from Taobao, containing 130 million real-world user interactions used in the experiments of this paper. The dataset is publicly available at \textcolor[rgb]{0.4, 0.5, 0.8}{\url{https://huggingface.co/datasets/PI2I/PI2I}}, which could serve as a valuable benchmark for the research community.
\end{abstract}



\maketitle

\section{Introduction}

\begin{figure}[t]
    \centering
    \includegraphics[width=0.5\textwidth]{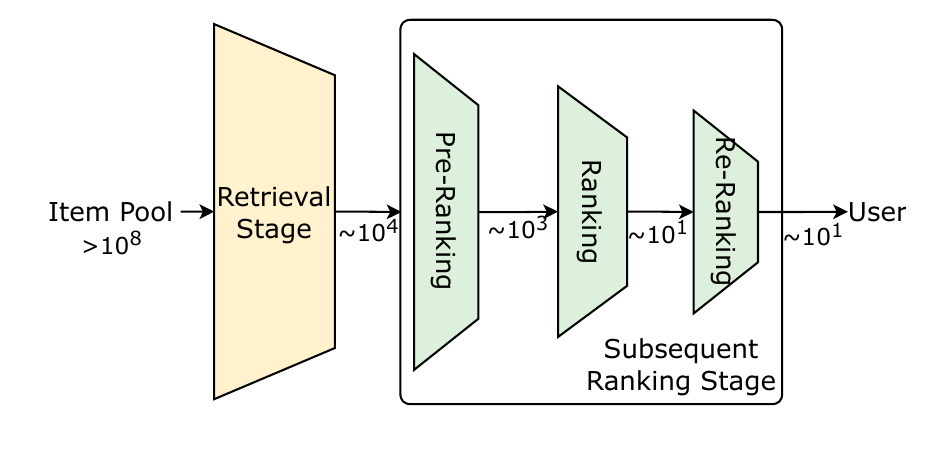}
    \vspace{-0.7cm}
    \caption{A general multi-stage architecture in modern recommender systems.}
    \label{fig:modernRecomender}
    \vspace{-0.3cm}
\end{figure}

In an era characterized by information overload, recommender systems (RS)~\cite{shani2011evaluating} have become crucial for online services like e-commerce~\cite{xie2022decoupled} and social media platforms~\cite{tang2022knowledge}. To manage vast content and meet time constraints, industries often use multi-stage cascade ranking systems. These systems primarily consist of two stages: retrieval~\cite{huang2024comprehensive} and ranking~\cite{pei2019personalized}. The retrieval, as shown in Figure~\ref{fig:modernRecomender}, rapidly narrows down a large pool of candidates to a smaller, more relevant set with an emphasis on scalability and low computational complexity. Following retrieval, the pre-ranking and ranking stages occur, where more detailed scoring and refinement processes build on the initial filtering done by the retrieval stage.

As the first gateway for user-item filtering in recommender systems, the retrieval stage is crucial for optimizing recommendation quality to preserve the most precious items. However, due to large candidate pools and strict time constraints, many advanced models are impractical for industrial use during this phase. Current retrieval methods are primarily classified into collaborative filtering~(CF) techniques~\cite{koren2021advances}, neural network approaches~\cite{cen2020controllable,manotumruksa2017deep,sun2019bert4rec} and graph-based methods~\cite{he2020lightgcn,chang2021sequential}. CF typically leverages user behavior data to statistically or vectorially compute item similarity. The advantages of this approach include both high performance and excellent interpretability. However, it has the drawback of \textbf{\romannumeral1)} lacking user personalization information in the offline truncation\footnote{Truncation in the retrieval stage caps the most relevant items (e.g., Top-1,000) for each historical item to balance computational cost and relevance before further ranking.} process, and \textbf{\romannumeral2) }consistent truncation size may also lead to the premature filtering of a large number of products that users might be interested in. Two-Tower neural models~\cite{huang2013learning,covington2016deep,li2019multi,cen2020controllable} separate user and item modeling for a balance of efficiency and effectiveness, utilizing deep learning but relying on Approximate Nearest Neighbors (ANN) for online inference. \textbf{\romannumeral3) }The overdue user-item crossing limits their ability to fully capture user-item interactions.
In recent times, Graph Neural Networks (GNNs)~\cite{wang2018billion,he2020lightgcn,wang2019neural} have been employed for different tasks to grasp detailed user-item interactions. Nonetheless, the complexity makes it quite challenging to implement GNNs in large-scale industrial contexts. Therefore, the development of a retrieval model that \textbf{meets tight time criteria} while adeptly \textbf{capturing complex user-item interactions} remains a formidable challenge.

Towards this end, we propose an innovative two-stage framework, \textbf{PI2I} for \textbf{P}ersonalized \textbf{I}tem-to-\textbf{I}tem collaborative filtering that balances scalability with model complexity for efficient user-item interaction capture. PI2I combines the strengths of collaborative filtering (CF) and advanced interaction models, optimizing both speed and precision in the retrieval stage. The whole process can be divided into the \textbf{I}ndexer \textbf{B}uilding \textbf{S}tage (IBS) and \textbf{P}ersonalized \textbf{R}etrieval \textbf{S}tage (PRS).
In the IBS, PI2I creates the initial item pool, strategically relaxing truncation thresholds to preserve a high Hit Rate while ensuring computational efficiency. This maximizes potential recommendations and provides a solid foundation for the next stage.
As for the subsequent PRS, it employs an interactive model for more delicate scoring, surpassing simple inner product calculations to better capture complex user-item interactions. This interactive approach overcomes traditional limitations and significantly boosts retrieval performance. Additionally, we propose to construct negative samples using a Trigger-Target\footnote{In this paper, the trigger can be regarded as the interacted item of each user.} relationship, representing an improvement over traditional Two-Tower negative sampling techniques. The proposed sampling strategy improves alignment between offline training and online inference, thereby enhancing model consistency and effectiveness. 

Our PI2I is not only a refinement of existing CF methods but also a strategic advancement that shows considerable performance improvements in offline experiments. It closes the gap with, and occasionally surpasses, mainstream Two-Tower models. 

In summary, the main contributions are as follows:
\begin{enumerate}
    \item We propose a two-stage retrieval framework named PI2I. \textbf{\romannumeral 1)} Initially, a CF method is employed to delineate the retrieval item pool. This approach relaxes the truncation threshold and retains as much of the Hitrate of the candidate set as possible. \textbf{\romannumeral 2)} Subsequently, the negative samples are constructed using the Trigger-Target relationship. Compared to the Two-Tower negative sampling, this ensures consistency between offline training sampling and online inference. \textbf{\romannumeral 3)} Finally, an interactive model is utilized for retrieval scoring, surpassing the limitations of inner product calculations, thereby enhancing the model's capability.
    \item This framework represents a personalized enhancement of the CF method. Offline experiments demonstrate that PI2I significantly outperforms CF and can exceed the performance of mainstream Two-Tower models.
    \item We have implemented it for retrieval on Taobao's homepage "Guess You Like" recommendations, achieving a 1.05\% online transaction improvement. Additionally, we plan to open-source a large-scale real retrieval dataset from Taobao.
\end{enumerate}

\section{Related Work}

\subsection{Collaborative Filtering}
Collaborative filtering (CF)~\cite{koren2021advances} is a vital technique utilized during the retrieval phase of recommender systems, emphasizing the identification of similarities through collated feedback. Initial CF methods~\cite{herlocker2002empirical,bell2007improved} concentrated on finding resemblances among users or items. In particular, user-based CF~\cite{wang2006unifying,zhao2010user} often relies on metrics like cosine similarity or the Pearson correlation coefficient, which are derived from user evaluations within a user-item rating matrix. In contrast, item-based CF~\cite{sarwar2001item,linden2003amazon} evaluates the similarity between items by analyzing user interaction patterns with those items. By thoroughly examining user-item connections, Swing~\cite{yang2020large} evaluates the bond between two items based on shared interactions. Subsequently, Matrix Factorization-based CF~\cite{mnih2007probabilistic,koren2009matrix,manotumruksa2017deep} emerged to proficiently represent both users and items within a latent factor space. These techniques predict user-item interactions, such as ratings, by calculating the dot product of their latent factors. 
\subsection{Neural Network Methods}
To maintain a balance between predictive accuracy and efficient inference, the Two-Tower method~\cite{yang2020mixed}, which segregates the computation of user-side and item-side features, stands as a leading approach in the retrieval phase. This technique capitalizes on the unique attributes of users and items to enhance retrieval performance, thus improving the system's overall effectiveness. For instance, DSSM~\cite{huang2013learning} transforms users and items into a shared dimensional semantic space. By maximizing the cosine similarity between their semantic vectors, it trains an implicit semantic model to facilitate retrieval tasks.
YouTubeDNN~\cite{covington2016deep} applies mean pooling to handle sequential features on the user side, and subsequently utilizes several fully connected layers to derive representations of user features.
Subsequently, certain researchers started investigating multi-interest models within retrieval algorithms. MIND~\cite{li2019multi} and ComiRec~\cite{cen2020controllable} employ capsule networks or attention networks to recognize users' varied interests. These methods facilitate retrieval, yielding a broader range of results by leveraging the identified interests.
Although the aforementioned approaches have achieved certain success through online experiments, they still depend on the two-tower architecture. During the training process, there is a lack of interaction between user features and item features, which limits the effectiveness of these methods.

\subsection{Graph-based Methods}
Early graph-based retrieval methods~\cite{grbovic2018real,barkan2016item2vec,mikolov2013efficient} mainly implement this by extending the principles of word embeddings to recommendation systems. For example, EGES [28] incorporates supplementary data (such as user demographics or item characteristics) into embeddings, enhancing the model’s grasp of context. Subsequently, researchers have begun implementing graph neural networks~\cite{fan2019graph,wu2022graph,wu2020comprehensive,zhou2020graph} in recommendation systems. NGCF~\cite{wang2019neural} efficiently utilizes higher-order neighborhood data via a multi-layer neural network, allowing the model to grasp complex user-item interactions. In contrast, LightGCN~\cite{he2020lightgcn} streamlines the collaborative filtering process by eliminating unnecessary elements from conventional graph-based methods, concentrating only on core neighborhood aggregation to boost computational efficiency without sacrificing performance. Although GNN-based retrieval methods have been notably successful in uncovering complex patterns within user-item interactions, enhancing their efficiency for large-scale retrieval remains a considerable challenge.

\begin{figure*}[t]
    \centering
    \includegraphics[width=1.0\textwidth]{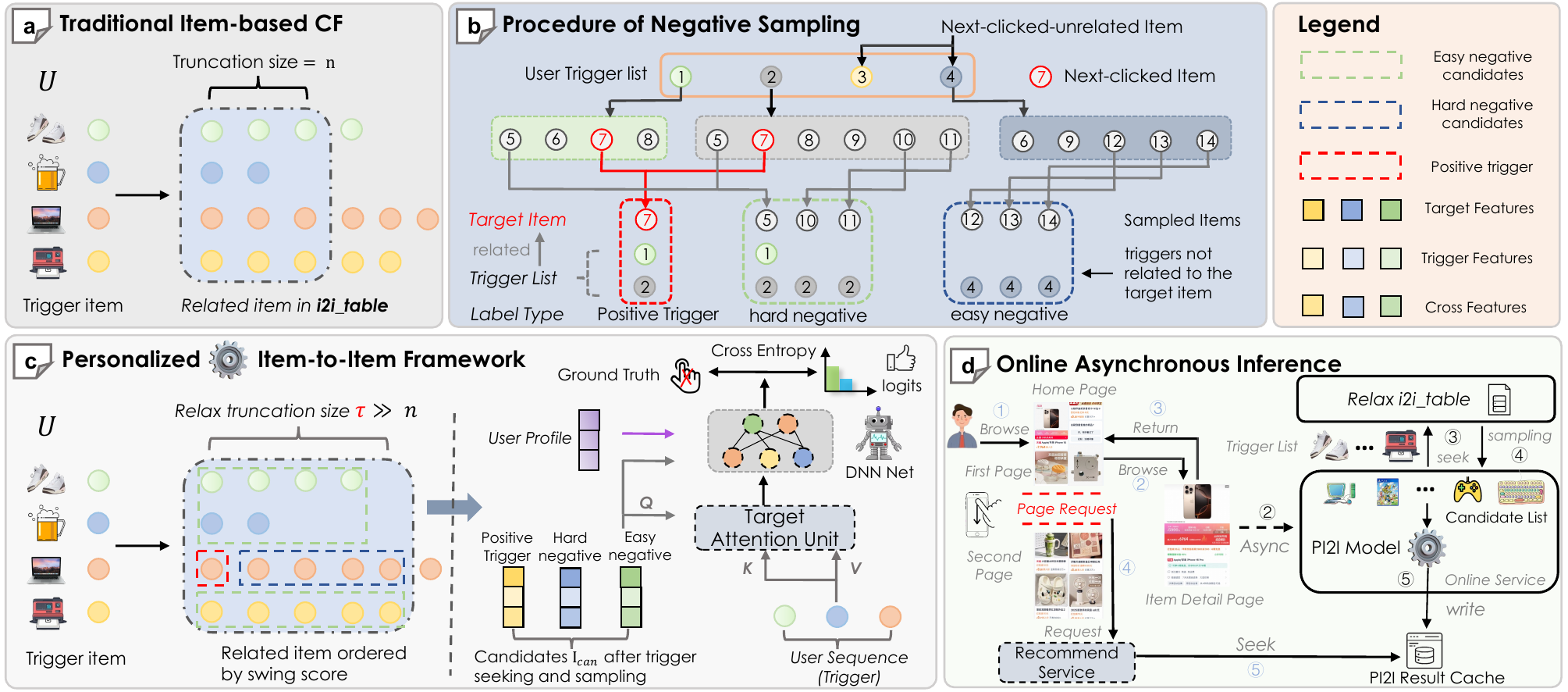}
    \vspace{-0.4cm}
    \caption{(a) Illustration of traditional I2I methods. (b) Illustration of the trigger-target sampling strategy. (c) The overall framework of PI2I \raisebox{-0.07cm}{\includegraphics[height=0.14in, width=0.14in]{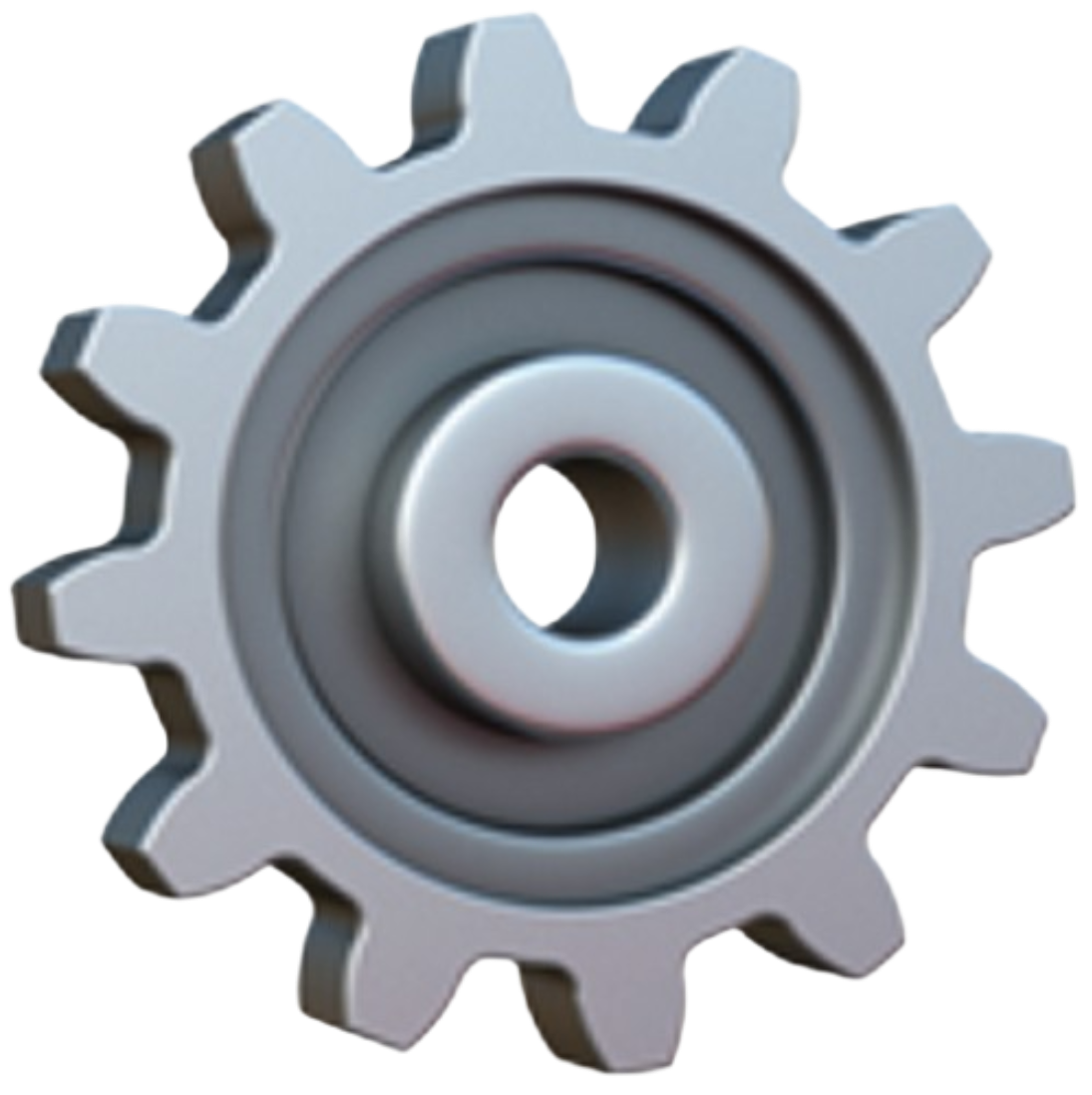}}. (d) Procedure of the PI2I Online Asynchronous Inference.}
    \vspace{-0.3cm}
    \label{fig:method}
\end{figure*}
\section{Methodology}

In this section, we will present in detail the Personalized Item-to-Item (PI2I) recommender system shown in Figure~\ref{fig:method}, including the Indexer Building Stage (IBS) in Section~\ref{IBS} and the Personalized Retrieval Stage (PRS) in Section~\ref{PRS}.

\subsection{Indexer Building Stage}
\label{IBS}
IBS stage is designed to efficiently retrieve tens of thousands of products from a vast inventory comprising billions of items $\mathcal{I}$ for each user $u\in \mathcal{U}$. Its primary objective is to construct an item-to-item table (i2i\_table) to reduce the scale of product listings while truncating to retain only those highly relevant $\mathcal{T}$ items for each $\mathrm{item}_i\in \mathcal{I}$ in the user behavior sequence:
\begin{equation}
\begin{aligned}
  \underbrace{\rm {\textbf{i2i\_table}}}_{\rm{Built\ by\ IBS}} : \underbrace{\rm {item_i}}_{\rm{Trigger}}  \xrightarrow[]{Mapping} \underbrace{\rm {[item^i_1,...,item^i_{\mathcal{T}}]}}_{\rm{Target (related\ items)}},
 \label{equ_1}
\end{aligned}
\end{equation}
where $[\mathrm{item}^i_1,...,\mathrm{item}^i_{\mathcal{T}}]$ are those $\mathcal{T}$ truncated items with the highest relevant scores with $\mathrm{item_i}$ and $\mathrm{item_i}\notin [\mathrm{item}^i_1,...,\mathrm{item}^i_{\mathcal{T}}]$. The index item $\mathrm{item}_i$ represents the \textbf{trigger}, while any of the related items from $\mathrm{item}^i_1$ to $\mathrm{item}^i_{\mathcal{T}}$ serve as the \textbf{target}. Together, they form the "Trigger-Target" relationship in the following section.

To accomplish the construction of i2i\_table, we employ an Item-to-Item Collaborative Filtering (item-based CF, or I2I) indexing approach, which leverages co-occurrence patterns to compute the similarity between items. Specifically, Swing~\cite{yang2020largescaleproductgraph} is selected for filtering in IBS due to its effectiveness in mitigating the noisy information, thereby enhancing the robustness of the predicted relationships. 
Formally, the definition of swing score is given below: 
\begin{equation}
s(i, j)=\sum_{u \in U_i \cap U_j} \sum_{v \in U_i \cap U_j} \frac{1}{\alpha+\left|I_u \bigcap I_v\right|},
\end{equation}
where $U_i$ and $U_j$ denote the set of users who have clicked item $i$ and $j$, separately. Similarly, $I_u$ and $I_v$ represent the set of items clicked by $u$ and $v$. $\alpha$ is a smoothing coefficient that avoids numerical instability caused by a small denominator. Additionally, to prevent the much influence of active users on $s(i,j)$, we apply another user weighting factor like the Adamic/Adar algorithm to penalize active users. The more items a user clicks, the smaller the weight it gets:
\begin{equation}
    s(i,j)=\sum_{u \in U_i \cap U_j} \sum_{v \in U_i \cap U_j} w_u\cdot w_v\cdot\frac{1}{\alpha+\left|I_u \bigcap I_v\right|},
\end{equation}
where
\begin{equation}
    w_u=\frac{1}{\sqrt{|I_u|}},\ w_v=\frac{1}{\sqrt{|I_v|}}.
\end{equation}
 
Ranking and truncating based on the swing score $s(i, j)$ in descending order yields the i2i\_table containing a set of highly related items for each item, as shown in Equation~\ref{equ_1}. Traditional item-based CF typically selects a small truncation size (e.g., 50) to mitigate online latency. Meanwhile, all the trigger items share a uniform truncation size. However, several more popular items are likely to be clicked with greater frequency, necessitating a larger truncation size to offer a broader array of comparable items.
To mitigate this issue, IBS adopts a substantially larger truncation size $\mathcal{T}$ to enhance the recall of the candidate set by including a broader range of relevant items. The differences between traditional I2I methods and PI2I are illustrated in Figure \ref{fig:method} (a) and (c).

The selection of a proper truncation size $\mathcal{T}$ is of great significance to balance the retrieval performance and computational resources. The Hit Rate (HR), which calculates the average proportion of final retrieved items $I_{K}$ with Top\_K scores that overlap with the total $N$ user click samples $I_{click}$ during this session:
\begin{equation}
\label{HR}
\text { HR@K }=\frac{1}{N} \sum_{m=1}^N \frac{I_{K} \cap I_{click}}{I_{click}}.
\end{equation}
It serves as the primary metric for determining the truncation size $\mathcal{T}$. Increasing the truncation size $\mathcal{T}$ enlarges the candidate pool, thereby enhancing the HR of relevant items. However, a much larger $\mathcal{T}$ also elevates the computational load (e.g., GPU utilization). 

In Section~\ref{parameter study}, we empirically identify the operational optimum truncation size $\mathcal{T}$ that maximizes recommendation quality within computational budget limitations. This derived truncation parameter subsequently governs the index table construction, ensuring alignment between statistical effectiveness and system efficiency in production environments.

\subsection{Personalized Retrieval Stage}
\label{PRS}
Assume that a user behavior log is defined as the historical items the user has clicked $U=[ item_1,item_2,...,item_n]$. After clicking on the \( n_{th} \) item $item_n$, the goal is to predict the $(n+1)_{th}$ item $item_{n+1}$. Traditional item-based CF would use all the clicked items $U$ as the trigger list to simply retrieve the final recall results from the i2i\_table. Nevertheless, such methods cannot differentiate the importance of each trigger, and the prediction with mere i2i\_table cannot directly perform end-to-end optimization of online performance. In contrast, PI2I extends it to the next-item prediction, optimizing the scoring beyond the construction of i2i\_table. With candidate items $\mathcal{I}_{\mathrm{can}}$ from i2i\_table with an intelligent trigger-target sampling strategy in Section~\ref{Trigger-Target Sampling}, PI2I subsequently performs the personalized retrieval by meticulously scoring items and selecting the Top-K items with the highest logits as the final retrieval results:
\begin{equation}
    \mathcal{F}, U, c_i \xrightarrow[]{\Theta} S_i, c_i \in \mathcal{I}_{\rm can},
\end{equation}
\begin{equation}
    |\rm item_{n+1}|=TopK(\{S_0,S_1..,S_{\mathcal{T}}\}),
\end{equation}
where $\Theta$ is the parameter of the score function in PRS to predict the clicked probability $S$ of the candidates in $\mathcal{I}_{\rm can}$ based on $U$ and other features $\mathcal{F}$ detailed in Section~\ref{Light and Efficient Model}. Items with the highest Top-K scores in $\mathcal{I}_{\rm can}$ would be chosen as the final retrieval results.

Through optimizing the scoring model $\Theta$, different triggers for the same user could reflect varying levels of importance, and the same trigger may have different impacts on different users, thus enabling personalization and allowing for direct end-to-end optimization of online target outcomes.

\subsubsection{Trigger-Target Sampling}
\label{Trigger-Target Sampling}
Conventional negative sampling methods in retrieval models include in-batch negative sampling and full random negative sampling. However, in PI2I, the candidate set is circumscribed to the targets triggered by clicks during IBS. As a result, not all items would be scored, and traditional negative sampling is no longer applicable.

During training, we employ a unique trigger-target index relationship for sampling shown in Figure \ref{fig:method} (b) and Algorithm~\ref{alg:Random}: (\textbf{\romannumeral1}): Treat the next-clicked item $\rm {item_{n+1}}$ of the user as positive item. (\textbf{\romannumeral2}): Only historical items in $U$ which are related to the positive item in i2i\_table are included in the positive trigger. (\textbf{\romannumeral3}): For those positive-triggered items, their target items from i2i\_table are randomly sampled as hard negative samples. (\textbf{\romannumeral4}): Easy negative items are sampled from the target items of those clicked but not triggered histories.

\begin{algorithm}[htb]
\caption{Trigger-target negative sampling strategy.}
\label{alg:Random}
\centering
\begin{algorithmic}[1]
    \Function{Sample}{$items,\ hard=\mathrm{TRUE}$}
        \State $Negative\_items \gets []$
        \If{hard}
            \State $Negative\_items \gets \text{Sample\ from\ }items \text{ with a higher } \sigma$
        \Else
            \State $Negative\_items \gets \text{Sample\ from\ }items \text{\ with a lower } \sigma$
        \EndIf\\
        \Comment{$\sigma$ denotes the sampling probability of each item in $items$}
        \State \Return $Negative\_items$
    \EndFunction
    \\
    \State $Target\ Item \gets Next\text{-}clicked\ Item$ \\ \Comment{e.g., the mobile phone browsed by users in Figure~\ref{fig:method}}
    \State $hard\ negatives \gets []$
    \State $easy\ negatives \gets []$
    \ForAll{$trigger \in \{user\_trigger\_list\}$}
        \State $related\_items \gets i2i\_table(trigger)$
        \If{$Target\ Item\in related\_items$}
            \State $retrieval\ items\gets \Call{Sample}{related\_items,\ \mathrm{True}}  $
            \State $hard\ negatives \gets hard\ negatives \cup retrieval\ items $
        \Else
            \State $retrieval\ items \gets \Call{Sample}{related\_items,\ \mathrm{False}}  $
            \State $easy\ negatives \gets easy\ negatives \cup retrieval\ items $
        \EndIf
    \EndFor
    \State \Return $easy\ negatives,\ hard\ negatives$
\end{algorithmic}
\end{algorithm}

This sampling method inherently introduces the association between triggers and the target, ensuring that the candidate set during inference and the sampling set during training are perfectly aligned, thereby optimizing both offline and online processes. Additionally, since the target items can be associated with multiple triggers simultaneously, this data information can further enhance the model's scoring capability, which will be elaborated in Section~\ref{Ablation Study}. However, it should be noted that if the next-clicked item does not exist among the target items of any triggers, the sample will be discarded. In practical scenarios of Taobao, approximately 15\% of samples may not find such associations.

\subsubsection{Light and Efficient Model}
\label{Light and Efficient Model}
During online inference, the actual scoring volume is at the level of 100,000. PI2I needs to retain complex computational capabilities while maintaining highly efficient inference speed. Compared to the traditional approach, which takes several user behavioral sequences of different action types as input, PI2I utilizes only a single trigger list as the model sequence for efficiency. Moreover, in this stage, we employ target attention with complex interactions, including a cleverly pre-designed interaction between triggers and targets in terms of features.

The architecture of PI2I can be referred to in Figure~\ref{fig:method}(c), which incorporates both user features and item features. User features encompass the user profile as well as the sequence of actions undertaken by the user over the past days. The behavioral sequence serves as a trigger for predicting subsequent items. To improve the responsiveness to these triggers and to amplify their influence on the activation of user interests, we concatenate the trigger features, target features, and cross-interaction features for prediction:
\begin{equation}
Q=M L P\left(E_{\text {trigger }} \oplus E_{\text {target }} \oplus E_{\text {cross }}\right),
\end{equation}
where $\oplus$ denotes the concatenation operation between features. The trigger and target features include categorical side information regarding the items, such as seller ID and brand ID. The cross-interaction features include elements such as price differences, co-click occurrences, rankings within the indexing table, and other relevant metrics about the relationship between the trigger and target items.

The concatenated embeddings are fed into an MLP layer and then serve as the Query input to the Multi-Head Target Attention (MHA) module~\cite{vaswani2017attention}, where user behavior sequences act as the corresponding Key and Value.
\begin{equation}
h e a d_i=\frac{\operatorname{Softmax}\left(Q K_{\text {seq }}^T\right) V_{\text {seq }}^T}{\sqrt{d_K}},
\end{equation}

\begin{equation}
MHA=\left[head_1 \oplus head_2 \oplus \ldots \oplus head_n\right].
\end{equation}
After the Multi-Head Target Attention Module, the output vector is concatenated with user profile embedding and item embedding, and then fed into another MLP layer to get the final logits.
\begin{equation}
   logits = MLP\left(MHA \oplus E_{\text {profile}} \oplus E_{\text{trigger}} \oplus E_{\text{target}} \oplus {E_{\text{cross}}} \right).
\end{equation}

\subsubsection{Loss Function}
In terms of model objectives, PI2I needs to ensure that the score of positive samples in the candidate set is greater than that of negative examples.
Negative examples is denoted as $\mathcal{V}$ while the positive item is denoted as \(p\). The objective function is to optimize the following negative likelihood:\\
\begin{equation}
\mathcal{L}^p=\sum_{b \in B}-\log \frac{\exp logits_{p}}{\exp logits_{p}+\sum_{k \in \mathcal{V}} \exp logits_{k}},
\end{equation}
where $B$ denotes the batch data for training.

\subsubsection{Inference}
Figure~\ref{fig:method} (d) depicts the online inference phase. Once a user clicks on an item, the asynchronous inference services would be triggered and PI2I dynamically leverages all items within the user's behavior sequence as contextual triggers. These triggers initiate a comprehensive retrieval process that systematically fetches all associated candidate items from the i2i\_table constructed in IBS, forming the complete scoring space of PRS. The model subsequently computes ranking scores across this isomorphic candidate set, with the top-K highest-scoring candidates constituting the final deterministic recommendation set. Further analysis of the process and data related to online inference can be found in Section \ref{Online Deployment}.

\section{Experiments}

\subsection{Dataset}

\begin{itemize}[leftmargin=*]
    \item \textbf{KuaiRec}~\cite{gao2022kuairec} is a comprehensive, real-life dataset with high density, gathered from Kuaishou's online platform. It comprises millions of dense user interactions alongside extensive auxiliary information. After some processing by the authors, this dataset is relatively less sparse, and the number of interactions is relatively smaller than those in industrial settings.
    \item \textbf{Taobao-Rec} dataset was gathered from Taobao, an e-commerce recommender system that handles a billion-scale of users and items. It was randomly sampled from the Taobao Mobile App, which contains a user-item interaction matrix over 20 days. As a large-scale industrial dataset, it is extremely sparse and can reflect the characteristics of a real recommendation task. Furthermore, Table 1 shows both components of the dataset. \textit{Taobao-Large} can be applied in large-scale distributed training models and is more closely aligned with the real-world recommendation task. \textit{Taobao-Small} can be used to verify and test offline evaluation for recommendation methods. 

\end{itemize}

\begin{table}[ht]
\tabcolsep=4.5pt
\vspace{-0.0cm}
\caption{Statistics of the dataset.}
\vspace{-0.3cm}
\label{tab:dataset}
\begin{tabular}{@{}ccccc@{}}
\toprule
                      & \textbf{\#users} & \textbf{\#Items} & \textbf{\#Interactions} & \textbf{Sparsity} \\ \midrule
\textit{KuaiRec} & 7,176            & 10,728            & 12,530,806               & 83.7\%        \\
\textit{Taobao-Large}   & 705,699            & 20,402,247           & 130,828,023             & 99.9\%           \\
\textit{Taobao-Small}   & 36,535            & 2,897,861           & 6,434,021              & 99.9\%           \\ 
\bottomrule
\end{tabular}
\tabcolsep=7pt
\vspace{-0.5cm}
\end{table}

\subsection{Experimental Setup}

\subsubsection{Evaluation Metrics}

To assess the performance of the evaluated methods, the task is formulated as predicting a user’s next interaction. We employ the Hit Rate (HR) described in Equation~\ref{HR} as the primary evaluation metric, as the proposed method is specifically designed to enhance retrieval effectiveness in the candidate matching phase. To align with this objective, we prioritize evaluating HR at larger top-K thresholds. Specifically, we set K=4000 for large-scale datasets and K=500 for smaller datasets to comprehensively measure coverage across both popular and long-tail items in the retrieved candidate sets.

\begin{table*}[t]
\centering
\caption{The performance of evaluated methods on Taobao Homepage. We use \textbf{bold} font to denote the best model and \underline{underline} the next best-performing model.}
\label{table:overall_result}
\vspace{-0.3cm}
\renewcommand{\arraystretch}{1.0}
\small
\begin{tabular}{llccccccc}
\toprule[1.5pt]
 \textbf{Dataset} & \textbf{Method} & \textbf{Hit@100} & \textbf{Hit@200} & \textbf{Hit@500} & \textbf{Hit@1000} & \textbf{Hit@2000} & \textbf{Hit@3000} & \textbf{Hit@4000} \\
\midrule
\midrule
\multirow{6}{*}{Taobao-Large} 
    & DSSM         & 1.10\% & 1.53\% & 2.31\% & 3.13\% & 4.18\% & 4.95\% & 5.57\% \\
    & YoutubeDNN   & 1.17\% & 1.65\% & 2.51\% & 3.40\% & 4.58\% & 5.40\% & 6.06\% \\
    & SASRec       & \textbf{2.65\%} & \textbf{3.75\%} & \textbf{5.61\%} & \underline{7.39\%} & \underline{9.51\%} & \underline{10.92\%} & \underline{12.00\%} \\
    & Swing        & 0.83\% & 1.15\% & 1.72\% & 2.31\% & 3.09\% & 3.67\% & 4.12\% \\
    & CORE         & 1.18\% & 1.65\% & 2.50\% & 3.40\% & 4.60\% & 5.45\% & 6.15\% \\
    \rowcolor{lightblue} \cellcolor{white} &  PI2I       & \underline{1.55\%} & \underline{2.61\%} & \underline{5.03\%} & \textbf{8.11\%} & \textbf{12.63\%} & \textbf{16.07\%} & \textbf{18.86\%} \\
\midrule
\midrule
\multirow{9}{*}{Taobao-Small} 
    & DSSM         & 1.41\% & 1.69\% & 2.14\% & 2.51\% & 3.07\% & 3.48\% & 3.74\% \\
    & YoutubeDNN   & \underline{2.92\%} & 3.36\% & 3.93\% & 4.45\% & 5.09\% & 5.48\% & 5.81\% \\
    & SASRec       & 2.82\% & \underline{3.61\%} & \underline{4.61\%} & 5.31\% & 6.02\% & 6.51\% & 6.87\% \\
    & Swing        & 1.55\% & 2.07\% & 2.87\% & 3.56\% & 4.34\% & 4.92\% & 5.51\% \\
    & LightGCN     & 2.71\% & 3.28\% & 3.81\% & 4.43\% & 5.45\% & 6.47\% & \underline{7.84\%} \\
    & CORE         & \textbf{3.26\%} & \textbf{4.00\%} & \textbf{4.81\%} & \underline{5.53\%} & 6.34\% & 6.87\% & 7.25\% \\
    & DiffRec      & 2.84\% & 3.46\% & 4.57\% & \textbf{5.56\%} & \underline{6.48\%} & \underline{7.07\%} & 7.51\% \\
    & RecDCL       & 2.01\% & 2.64\% & 3.63\% & 4.67\% & 5.69\% & 6.37\% & 6.97\% \\
    \rowcolor{lightblue} \cellcolor{white} & PI2I         & 1.52\% & 2.14\% & 3.46\% & 5.08\% & \textbf{7.72\%} & \textbf{9.89\%} & \textbf{11.92\%} \\
\bottomrule[1.5pt]
\end{tabular}
\end{table*}

\subsubsection{Baselines}
For a complete evaluation of our method’s performance, we conduct comparisons with item-based Collaborative Filtering methods, embedding-based retrieval (EBR-based) methods, and graph neural network (GNN-based) methods.
\begin{enumerate}[leftmargin=*]
    \item \textbf{DSSM}~\cite{huang2013learning} maps user \& item into a low-dimensional semantic space, enabling efficient similarity calculation and matching through cosine similarity.
    \item \textbf{YoutubeDNN}~\cite{covington2016deep} aggregates the user historical behavior through average pooling of item embeddings to generate a unified user representation for recommendation candidate generation.
    \item \textbf{SASRec}~\cite{kang2018selfattentivesequentialrecommendation} leverages self-attention mechanisms to capture dynamic user behavior patterns and long-range dependencies in interaction sequences for next-item prediction.
    \item \textbf{Swing}~\cite{yang2020large} proposes that if multiple users purchase the same pair of items simultaneously, with relatively few other co-purchase behaviors, then the association between these items is more credible.
    \item \textbf{LightGCN}~\cite{he2020lightgcn} is a simplified graph convolutional network-based recommendation algorithm that enhances collaborative filtering by efficiently learning both user and item embeddings through neighbor aggregation.
    \item \textbf{CORE}~\cite{hou2022coresimpleeffectivesessionbased} designs an encoder that uses a linear combination of item embeddings as the session embedding, ensuring that sessions and items reside in the same representation space.
    \item \textbf{DiffRec}~\cite{wang2023diffusionrecommendermodel} proposes a diffusion model-based method to generate user interaction, which adds noise to the user's historical interactions and then infers the interaction probabilities through a denoising process.
    \item \textbf{RecDCL}~\cite{Zhang_2024} is a contrastive learning based Self-supervised learning method for enhancing collaborative filtering. It combines batch contrastive learning and feature contrastive learning to optimize the unified distribution within users and items and enhance the robustness of these representations.
\end{enumerate}

\subsubsection{Implementation details}
For all experimental configurations, the embedding dimension and mini-batch size were fixed at 64 and 512 across all models, respectively. Model optimization was conducted using the Adam optimizer~\cite{adam2014method} with a fixed learning rate of 0.01. In GNN-based architectures, the graph neural network layer depth was set to 2, selected from {1,2,3} according to comprehensive validation. For transformer-based baselines, the number of transformer blocks is selected out of {1,2,3,4}. For fair comparison, we implemented representative methods under the same framework RecBole~\cite{recbole}. For GNN-based and SSL-based methods, the implementation on the large dataset results in Out-Of-Memory error, so we only compare them on KuaiRec and Taobao-Small Dataset. 

\begin{table}[hb]
    \centering
    \vspace{-0.3cm}
    \caption{The performance of evaluated methods on KuaiRec.}
    \label{table:kuai_result}
    \vspace{-0.3cm}
    \renewcommand{\arraystretch}{1.0}
    \setlength{\tabcolsep}{4pt}
    \resizebox{\columnwidth}{!}{
    \begin{tabular}{lcccccc}
        \toprule[1.5pt]
        \textbf{Method} & \textbf{Hit@20} & \textbf{Hit@50} & \textbf{Hit@100} & \textbf{Hit@200} & \textbf{Hit@500} \\
        \midrule
        \midrule
        DSSM & 1.65\% & 3.50\% & 5.91\% & 9.97\% & 19.03\% \\ 
        YoutubeDNN & 1.90\% & 4.31\% & 7.42\% & 12.35\% & 23.86\% \\ 
        SASRec & 0.94\% & 2.31\% & 5.22\% & 11.94\% & \underline{32.41\%} \\ 
        Swing & 0.28\% & 0.51\% & 1.44\% & 2.53\% & 5.59\% \\  
        LightGCN & 1.35\% & \underline{4.44\%} & \underline{7.56\%} & \underline{13.95\%} & 20.45\% \\ 
        CORE & 1.89\% & 4.09\% & 6.92\% & 11.79\% & 24.36\% \\
        DiffRec & 1.93\% & 3.90\% & 6.74\% & 10.92\% & 18.43\% \\ 
        RecDCL & \underline{2.13\%} & 4.27\% & 7.18\% & 11.77\% & 22.00\% \\ 
        \rowcolor{lightblue} PI2I & \textbf{2.33\%} & \textbf{5.41\%} & \textbf{9.96\%} & \textbf{17.52\%} & \textbf{34.26\%} \\
        \bottomrule[1.5pt]
    \end{tabular}
    }
    \vspace{-0.6cm}
\end{table}

\subsection{Overall Performance}
In this subsection, we compare our proposed PI2I with the state-of-the-art and classic models on three datasets. The comparison results on the Taobao Dataset are presented in Table~\ref{table:overall_result}, and the comparison results on the KuaiRec Dataset~\cite{gao2022kuairec} are illustrated in Table~\ref{table:kuai_result}. From the results, we have the following observations:

On the dense KuaiRec dataset, the proposed PI2I model demonstrates comprehensive superiority, significantly outperforming state-of-the-art baselines across all top-K HR metrics from HR@20 to HR@500, as evidenced by the benchmark results in Table~\ref{table:overall_result}. Meanwhile, we observe GNN-based methods such as LightGCN perform better than EBR-based methods on HR@top50-400, while EBR-based methods such as SASRec and CORE perform better than GNN-based methods on  HR@top 500.

On the sparse Taobao dataset, the PI2I model demonstrates a clear performance trade-off: it significantly outperforms the baseline in HR at larger top-K thresholds (e.g., more than 40\% improvements on HR@4000 compared to the second-best method), but fails to surpass the baseline at smaller ranges (e.g., HR@20). This disparity arises from its hybrid architecture, which integrates collaborative filtering (CF) and embedding-based retrieval (EBR) strategies, enabling it to balance the diversity of CF-based methods and personalization effectively.
As a retrieval-stage model in industrial settings, PI2I is specifically designed to generate candidate sets containing thousands of items (K = 3000–4000) for the subsequent ranking phases. Therefore, its optimization aligns closely with maximizing large-range HR metrics. While its performance at smaller top-K values (e.g., 20) does not exceed that of pure EBR baselines, this limitation is compensated for by downstream ranking processes that enhance precision within the broader candidate pool.

\begin{figure}[ht] 
    \centering
    \includegraphics[width=1.0\linewidth]{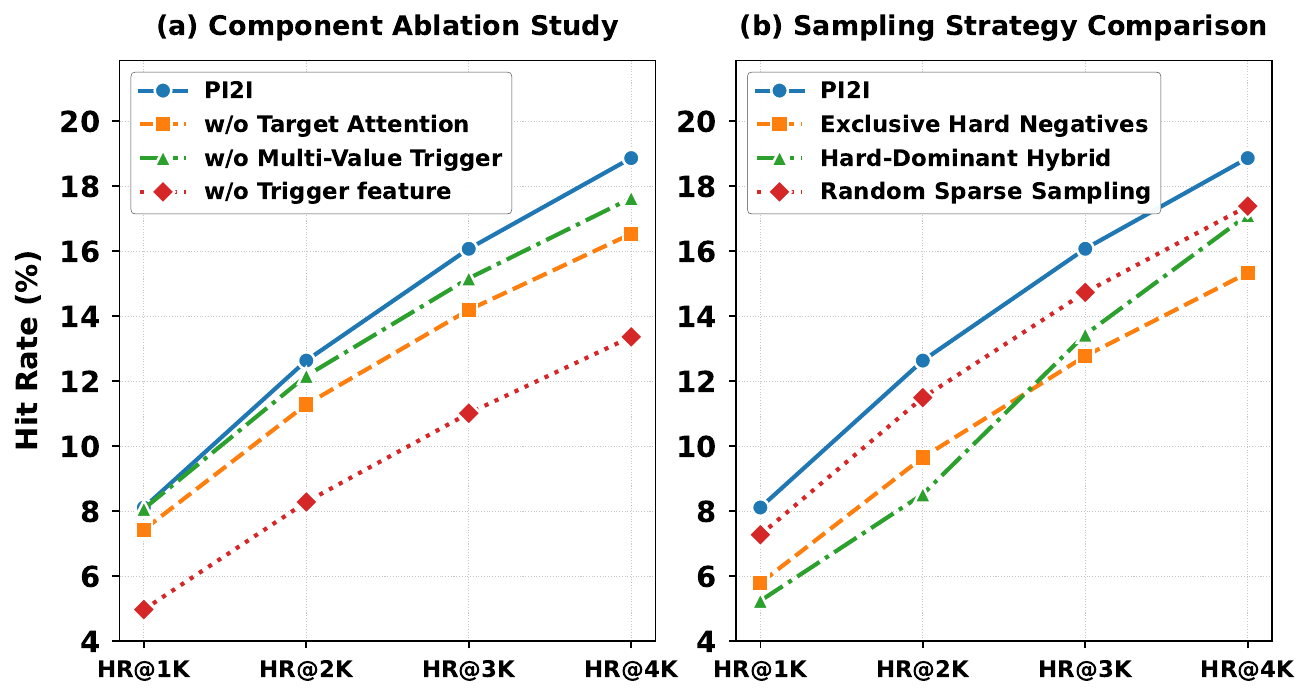} 
    \vspace{-0.6cm}
    \caption{Ablation study.}
    \label{fig:ablation_study}
    \vspace{-0.4cm}
\end{figure}

\subsection{Ablation Study}
\label{Ablation Study}
The first ablation study is focused on three key components within the PRS stage: (1) Target Attention Ablation (\textbf{w/o Target Attention}), which eliminates the Target Attention (TA) module and substitutes it with standard self-attention mechanisms; (2) Multi-Value Trigger Simplification (\textbf{w/o Multi-Value Trigger feature}), where we discard the multi-value trigger representation and instead employ random selection from available triggers to maintain a single-value trigger feature; (3) Full Trigger Removal (\textbf{w/o Trigger feature}), which completely eliminates the trigger feature from the model architecture. We conducted ablation studies on the Taobao-Small dataset, with results shown in Figure~\ref{fig:ablation_study}. The key observations from it include:

\begin{itemize}[leftmargin=*]
    \item \textbf{TA outperforms Self-Attention in the retrieval model}. Replacing TA with Self-Attention results in a decrease in HR, highlighting the effectiveness of target interaction in the recommendation retrieval model.
    \item \textbf{The Multi-Value Trigger feature enhances the scoring model}. In contrast, the Single-Value Trigger approach, which randomly selects one trigger item per target item, reduces information gain and leads to lower performance. Furthermore, removing all triggers significantly degrades performance compared to the other setups.
\end{itemize}

The second ablation study is to investigate the impact of negative sample composition ratios on model performance, as illustrated in Figure~\ref{fig:ablation_study}(b). Three distinct sampling strategies were systematically compared: 
\begin{itemize}[leftmargin=*]
    \item \textbf{Exclusive Hard Negatives}: All negative samples (N=20 per positive) were drawn from the same triggers of the target item.
    \item \textbf{Hard-Dominant Hybrid}: Maintained baseline hard negative sample count (N=20) with only 10 easy negatives (1/8 of baseline easy negatives).
    \item \textbf{Random Sparse Sampling}: Employed random selection of both negative types at 50\% baseline density (N=50), preserving original hard/easy ratio but reducing absolute quantities.
\end{itemize}

Based on the figure, we have the following conclusions: \textbf{Both hard negatives and easy negatives are crucial elements in the sampling procedure}. Relying exclusively on hard negative results in suboptimal outcomes, as these items differ from the prediction scoring space. Furthermore, a reduction in the quantity of negative samples may also contribute to a decline in model performance.

\begin{figure}[ht] 
    \centering
    \includegraphics[width=0.38\textwidth]{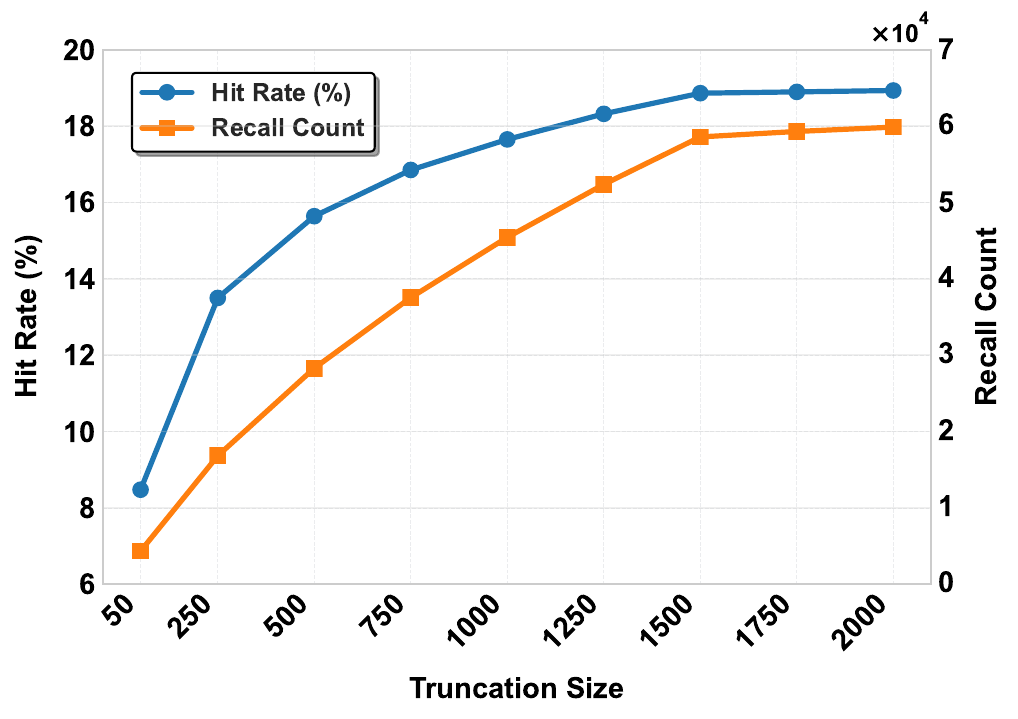} 
    \vspace{-0.2cm}
    \caption{Truncation Parameter Trade-off.}
    \label{fig:truncation_hr}
    \vspace{-0.4cm}
\end{figure}
\subsection{Parameter Study}
\label{parameter study}
In this subsection, we study the impact of the key parameter truncation size $\mathcal{T}$ in the IBS stage. With the increase in the truncation size of the indexer table, the slope of the HR becomes slower. As a trade-off between performance and system efficiency, we need to select a proper parameter of truncation size. A much longer truncation size may result in delayed processing of user requests after the deployment of the retrieval model. The relationship of truncation size and HR is shown in Figure~\ref {fig:truncation_hr}. 
By three key observations from the figure, $\mathcal{T}=1250$ is identified as the optimal truncation size for the IBS phase.

\textbf{Performance Saturation}. When the truncation size exceeds 1250, the HR exhibits marginal improvements of less than 0.5\% per 250-unit increment, indicating diminishing returns. 

\textbf{Cost-Efficiency Threshold}. Below $\mathcal{T}=1500$, total recall count as well as retrieval operations decrease exponentially (46\% reduction when halving $\mathcal{T}$), while HR degrades linearly (15\% relative drop). Setting $\mathcal{T}$ = 1250 reduces retrieval operations by 12.5\% compared to $\mathcal{T}$ = 1500, with only a 3.4\% sacrifice in HR.

\textbf{Operational Stability}. The $\mathcal{T}=1250$ configuration preserves 97\% of maximum HR while keeping latency and resource utilization within safe limits at Taobao platform. This balance ensures sustainable system performance without compromising recommendation quality.

\subsection{Case Study}
\begin{figure}[t]
    \centering
    \includegraphics[width=0.4\textwidth,height=5cm]{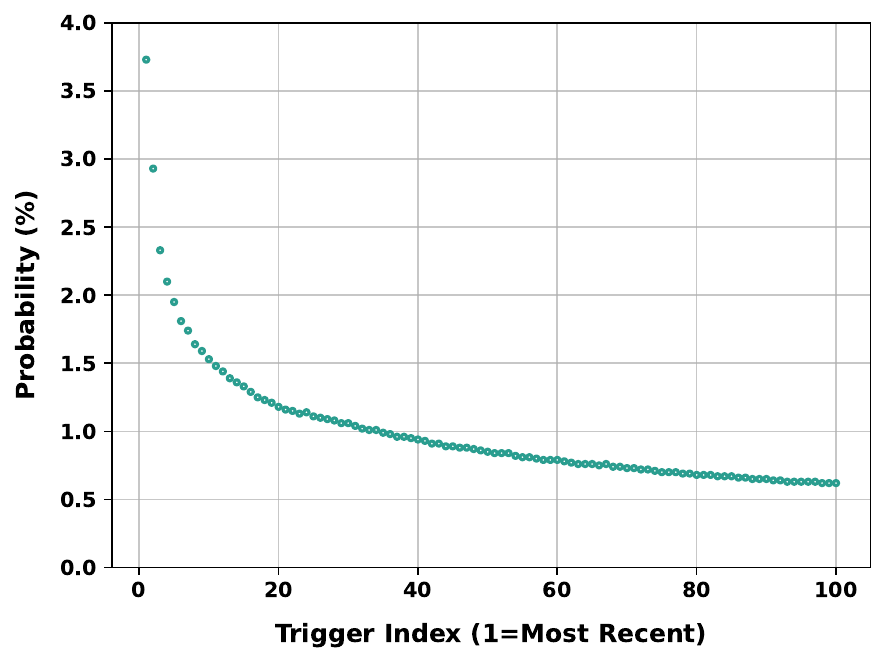} 
    \vspace{-0.2cm}
    \caption{Temporal Decay Pattern of Behavior Triggering.}
    \label{fig:trigger_trends}
    \vspace{-0.3cm}
\end{figure}

This subsection investigates the temporal decay characteristics of behavioral triggering mechanisms and demonstrates the personalized nature of trigger distributions among users. As evidenced in Figure~\ref{fig:trigger_trends}, the observed exponential decay pattern reveals a strong inverse correlation between trigger probability and behavioral recency, where proximal interactions (index=1) exhibit 6.02× higher triggering likelihood than distant ones (index=100), confirming the critical role of temporal proximity in interaction modeling.

\begin{figure}[t]
    \centering
    \includegraphics[width=0.37\textwidth]{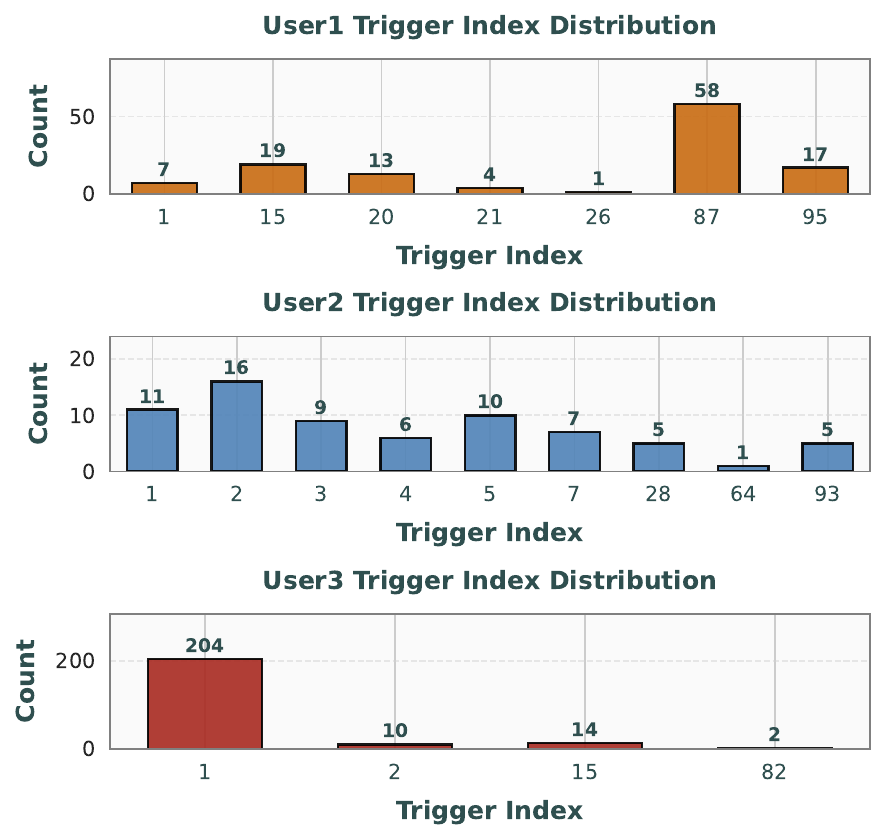} 
    \vspace{-0.2cm}
    \caption{Trigger Index Distribution Analysis.}
    \label{fig:trigger_distribution}
    \vspace{-0.2cm}
\end{figure}

However, an examination of results across different users reveals significant personalization in trigger distributions, as illustrated in Figure~\ref {fig:trigger_distribution}. We counted the clicks of 3 users over a period of time, with the vertical axis representing the number of clicks and the horizontal axis representing the trigger index that triggered each click. This personalization suggests that different users may have varying trigger positions that influence their next-click items. For instance, a user who purchased colas a month ago may exhibit a tendency to re-buy them recently, resulting in a larger trigger index value within the 1-100 range. Conversely, another user with a strong recent interest in a variety of dresses would likely have a lower trigger index, closer to 1, reflecting her frequent clicks on dresses. The findings presented in Figure~\ref{fig:trigger_distribution} demonstrate that the PI2I model effectively captures these trends.

Furthermore, distinct users exhibit varying distributions of trigger indices, which underscores the efficacy of the PRS stage. Unlike traditional item-based collaborative filtering methods, which truncate the index at a uniform position, the model demonstrates the capability to adjust truncation in accordance with user profiles and their specific behaviors.
  
\subsection{Online Deployment}
\label{Online Deployment}

\subsubsection{Online Implementation}
As illustrated in Figure \ref{fig:method}(d), we use the Nearline-Deployment~\cite{li2021truncationfreematchingdisplayadvertising} to address the scalability challenges. 
When a user clicks a product detail page, an asynchronous PI2I inference request is triggered. Tens of thousands of candidates will be retrieved by the IBS stage. In order to obtain pertinent items, the PRS stage employs a scoring model to narrow down the pool of candidates. The inference writes PI2I results into the cache table in 200ms, ensuring no impact on the main recommendation service's performance. The recommendation service could search the cache table and get the matching results asynchronously when browsing and requesting.
During subsequent browsing, the main recommendation service cached PI2I results in 5ms with over 95\% hit rate.
PI2I’s design ensures real-time application of user clicks to the next recommendation results.

\subsubsection{Online A/B Tests}
We carried out a 15-day randomized trial involving 5\% of Taobao's traffic, focusing solely on PI2I recall as the differentiating factor between the control group (using base retrieval methods) and the test group (base retrieval plus PI2I). The base retrieval techniques included Swing and DM++ (deep match with industrial optimization enhancements). Our online system monitored HR, CTR, TPM (Transactions Per Mille), and PVR (proportion of exposure) to assess performance. The results, detailed in Table 4, demonstrate that PI2I outperforms the other methods, boosting user transaction intent and proving advantageous for large-scale services. On Taobao's homepage, which receives hundreds of millions of views daily, PI2I achieved an improvement of +0.8\% in HR, contributing to a notable increase in product transactions by +1.05\%. 

\begin{table}[h]
    \caption{online HR, CTR, TPM and PVR performance. 
    }
    \setlength{\tabcolsep}{2pt}
    \centering
    \begin{tabular}{cccccccc}
    \toprule
    & \textbf{Methods} &  \textbf{Recall Size} & \textbf{HR} & \textbf{CTR}  & \textbf{TPM} & \textbf{PVR} \\ \hline
     & Swing & 5,225  & 6.30\% & base & base  & 10.59\% \\
     & DM++ & 4,399  & 8.96\% & +11.27\% & +57.30\% & 26.00\% \\
    \rowcolor{lightblue} & PI2I & 4,094 & \underline{\textbf{12.26\%}} &  \underline{\textbf{+51.27\%}} & \underline{\textbf{+116.09\%}}  & \underline{\textbf{ 31.70\%}} &   \\
    \bottomrule
    \end{tabular}
    \label{table:overall_result}
    \vspace{-2mm}
\end{table}

\section{Conclusion}
In this paper, we present a novel two-stage retrieval framework designed to enhance personalization in item-based collaborative filtering. Through our experimental setup, we have demonstrated the superior performance of our framework on both dense and sparse datasets. Additionally, we validated its effectiveness through long-term online A/B testing, leading to its deployment for all users in the "Guess You Like" section on the Taobao homepage. Looking ahead, we plan to further explore the information from the indexer table to enhance the performance of the online matching system.

\bibliographystyle{ACM-Reference-Format}
\balance
\bibliography{acmart}

\end{document}